**Concanavalin A-targeted mesoporous silica nanoparticles for infection treatment**

Marina Martínez-Carmona,[a,b] Isabel Izquierdo-Barba,[a,b] Montserrat Colilla,[a,b*]

María Vallet-Regí[a,b*]

[a] Dpto. Química Inorgánica y Bioinorgáni*ca* Universidad Complutense de Madrid. Instituto de Investigación Sanitaria Hospital 12 de Octubre i+12. Plaza Ramón y Cajal s/n, 28040 Madrid, Spain.

[b] CIBER de Bioingeniería, Biomateriales y Nanomedicina, CIBER-BBN, Madrid, Spain.

[*] Corresponding authors: Fax: +34 394 1786; Tel.: +34 91 394 1843; E-mail addresses: vallet@ucm.es (M. Vallet-Regí) and mcolilla@ucm.es (M. Colilla)

**Abstract**

The ability of bacteria to form biofilms hinders any conventional treatment against chronic infections and has serious socio-economic implications. In this sense, a nanocarrier capable of overcoming the barrier of the mucopolysaccharide matrix of the biofilm and releasing its loaded-antibiotic within would be desirable. Herein, a new nanosystem based on levofloxacin (LEVO)-loaded mesoporous silica nanoparticles (MSNs) decorated with lectin Concanavalin A (ConA) has been developed. The presence of ConA promotes its internalization into the biofilm matrix, which increases the antimicrobial efficacy of the antibiotic hosted within the mesopores. This nanodevice is envisioned as a promising alternative to conventional infection treatments by improving the antimicrobial efficacy and reducing side effects.

**Statement of Significance**

The present study is focused on finding an adequate therapeutic solution for the treatment of bone infection based on nanocarriers capable of overcoming the biofilm barrier by increasing the therapeutic efficacy of the loaded antibiotic. In this sense, we present a nanoantibiotic that increases the effectiveness of levofloxacin to destroy the biofilm formed by a model bacterium such as *E. coli*. used as a model. This work opens new lines of research in the treatment of chronic infections based on nanomedicines.

**1. Introduction**

Nowadays, antimicrobial resistance (AMR) constitutes a major threat of global health with serious socio-economic implications [1,2]. Therefore a post-antibiotic era is emerging to replace ineffective conventional antimicrobial treatments [3-5]. In this sense, the combination of AMR and bacterial biofilm formation leads to almost unmanageable infections [6]. A biofilm is a bacterial community in which bacteria are embedded into an extracellular matrix mainly composed of polysaccharides. It constitutes a natural mechanism of defense of the microorganisms against external aggressions, including antibiotics and immune system [7].

MSNs have recently entered the nanomedicine scenario due to their capability to host, protect and transport diverse drugs and locally release them once the target tissue is reached [8-10]. For further applications in bacterial infections, MSNs has proven to be a multifunctional and versatile solution, since they have advantages in all stages of combatting the infection including early detection, drug release, targeting bacteria or biofilm, antifouling surfaces and adjuvant capacity [11]. Specifically, once the biofilm has been formed, the use of these nanocarriers could be quite potent since their surface can be functionalized with targeting agents that increased the affinity towards biofilms and favor higher treatment efficiency [9-15].

Lectins such as ConA are glycoproteins that are present in a variety of organisms, and most of them are isolated from plant components [16]. They moreover have the ability to weakly bind glycans with high specificity to form glycoconjugates [17]. In actuality, ConA has been successfully used to design antitumour drug-loaded nanoparticles by selectively binding and internalizing in cancer cells overexpressing membrane glycans [18,19]. Biocompatibility was also demonstrated, as no significant cell death was observed after incubation with MC3T3-E1 (mouse preosteoblastic) cells at concentrations up to 144 µg/mL [19]. Since glycan-type polysaccharides are also present in the bacterial biofilm, we hypothesize that ConA could be used to target MSNs towards extracellular biofilm matrix. In fact, although the use of ConA for the treatment of planktonic bacteria is more widespread [20-22], its application once the biofilm is formed is very limited and there appears to be very few publications on the matter. And when uniquely present, ConA can be used for detection [23,24], having found a single publication in which the ConA is part of a nanosystem with treatment purposes [25]. In addition, this appears to be the initial demonstration of a ConA anchorage to the surface of nanoparticles that presents a significant antibiotic action, without the need for a loaded drug; although, it is true that the combined action of both ConA and a therapeutic would be thought to intensify the antimicrobial character of the system. Herein, we report the design of a new nanoantibiotic consisting of MSNs loaded with an antimicrobial agent (LEVO) and grafted in their outermost surface with ConA, which has been proved to selectively recognize and bind to certain glycans (Fig. 1).

## 2. Materials and Methods

### 2.1 Reagents

Tetraethylorthosilicate (TEOS, 98%), n-cetyltrimethylammo-nium bromide (CTAB, ≥ 99%), sodium hydroxide (NaOH, ≥ 98%), ammonium nitrate ($NH_4NO_3$, ≥ 98%), sodium carbonate ($Na_2CO_3$, ≥ 99,5%), hydrochloric acid (HCl, 37%), Rhodamine B isothiocyanate (RITC, ≥ 98%), (3-aminopropyl) triethoxysilane (APTES, ≥ 98%), N-(3-Dimethylaminopropyl)-N′-ethylcarbodiimide hydrochloride (EDC, ≥ 98%), N-Hydroxysulfosuccinimide sodium salt (sulfo-NHS, ≥ 98%), phosphate-buffered saline (PBS, 10x), phosphotungstic acid hydrate (PTA, reagent grade) and Concanavalin A from Canavalia ensiformis (Jack bean) (ConA, Type VI lyophilized powder) were purchased from Sigma-Aldrich (St. Louis, USA). 3-(Triethoxysiyl)propylsuccinic anhydride (SATES, 95%) was purchased from ABCR (Karlsruhe, Germany). All other chemicals were purchased from Panreac Química SLU (Castellar del Valles, Barcelona, Spain) inc: absolute ethanol, etc. All reagents were used as received without further purification. Ultrapure deionized water with resistivity of 18.2 MΩ was obtained using a Millipore Milli-Q plus system (Millipore S.A.S, Molsheim, France). Levofloxacin (LEVO, $C_{18}H_{20}FN_3O_4$, 98 %w) was purchased from Sigma Aldrich.

### 2.2 Characterization techniques

Powder X-Ray Diffraction (XRD) experiments were performed in a Philips X'Pert diffractometer equipped with Cu Kα radiation (wavelength 1.5406 Å) (Philips Electronics NV, Eindhoven, Netherlands). XRD patterns were collected in the 2ɵ range between 0.6° and 8° with a step size of 0.02° and counting time of 5 s per step. Thermogravimetric (TG measurements were performed in a Perkin Elmer Pyris Diamond TG/DTA (California, USA), with 5 °C $min^{-1}$ heating ramps, from room temperature (RT) to 600 ºC. Fourier transform infrared spectroscopy (FTIR) was carried out in a Nicolet (Thermo Fisher Scientific, Waltham, MA, USA) Nexus spectrometer equipped with a Goldengate attenuated total reflectance (ATR) accessory (Thermo Electron Scientific Instruments LLC, Madison, WI USA). Morphology, mesoestructural order and nanoparticles functionalization were studied by High Resolution Transmission Electron Microscopy (HRTEM) with a JEOL JEM JEM 1400 instrument, equipped with a CCD camera (JEOL Ltd., Tokyo, Japan). Sample preparation was performed by dispersing in distilled water and subsequent deposition onto carbon-coated copper grids. A 1% PTA solution (pH 7.0) was used as staining agent in order to visualize the organic coating around MSNs.

To determine the evolution of the size and surface charge of nanoparticles by dynamic light scattering (DLS) and zeta ($\zeta$)-potential measurements, respectively, a Zetasizer Nano ZS (Malvern Instruments, United Kingdom) equipped with a 633 nm "red" laser was used. DLS measurements were directly recorded in ethanolic colloidal suspensions. $\zeta$-potential measurements were recorded in aqueous colloidal suspensions. For this purpose, 1 mg of nanoparticles was added to 10 mL of solvent followed by 5 min of sonication to obtain a homogeneous suspension. In both cases, measurements were recorded by placing 1 mL of suspension (0.1 mg mL$^{-1}$) in DTS1070 disposable folded capillary cells (Malvern Instruments). The textural properties of the materials were determined by N$_2$ adsorption porosimetry by using a Micromeritics ASAP 2020 (Micromeritics Co., Norcross, USA). To perform the N$_2$ measurements, 20-30 mg of each sample was previously degassed under vacuum for 24 h at 40 ºC temperature. The surface area ($S_{BET}$) was determined using the Brunauer-Emmett-Teller (BET) method and the pore volume ($V_P$) was estimated from the amount of N$_2$ adsorbed at a relative pressure around 0.97. The pore size distribution between 0.5 and 40 nm was calculated from the adsorption branch of the isotherm by means of the Barrett-Joyner-Halenda (BJH) method. The mesopore size ($D_P$) was determined from the maximum of the pore size distribution curve.

**2.3 Synthesis of pure-silica MSNs (MSN)**

Bare MSNs, denoted as MSN, were synthesized by the modified Stöber method using TEOS as silica source in the presence of CTAB as structure directing agent. Briefly, 1 g of CTAB, 480 mL of $H_2O$ and 3.5 mL of NaOH (2 M) were added to a 1,000 mL round-bottom flask. The mixture was heated to 80 ºC and magnetically stirred at 600 rpm. When the reaction mixture was stabilized at 80 ºC, 5 mL of TEOS were added dropwise at 0.33 mL min$^{-1}$ rate. The white suspension obtained was stirred during further 2 h at 80 ºC. The nanoparticles were collected by centrifugation, washed twice with water and twice with ethanol and storage in an ethanol suspension.

For cellular internalization studies rhodamine-labeled MSN were synthesized. For this purpose, 1 mg FITC and 2.2 µL APTES were dissolved in 100 µL ethanol and left reacting for 2 h. Then the reaction mixture was added with the 5 mL of TEOS as previously described.

**2.4. Functionalization of MSN with carboxylic acid groups (MSN$_{SATES}$)**

With the aim of preferentially grafting carboxylic acid groups to the external surface of MSN. 500 mg of CTAB-containing MSN were placed in a three-neck round bottom flask and dried at 80 ºC under vacuum for 24 h. Then, 125 mL of dry toluene was added, and the flask was placed in an ultrasonic bath for several sonication cycles of 5 min until a good nanoparticles suspension was achieved. After that 300 µL of SATES were added, keeping the reaction under nitrogen atmosphere at 90 ºC for 24 h. The reaction mixture was centrifuged and washed three times with water and ethanol. The surfactant was removed by ionic exchange by soaking soaking 1g of nanoparticles in 500 mL of a $NH_4NO_3$ solution (10 mg mL$^{-1}$) in ethanol (95%) at 65 ºC overnight under magnetic stirring. The nanoparticles were collected by centrifugation, washed three times with ethanol and storage in an ethanol suspension.

**2.5. ConA grafting to MSN$_{SATES}$ (MSN$_{ConA}$)**

16 mg of MSN$_{SATES}$ were placed in a vial and suspended in 2 mL PBS pH 7.4 and subjected to several sonication cycles of 5 min until a good suspension was achieved. After that, 32 mg of EDC were added and the mixture was stirred at R.T. for 40 min. Then 14 mg of sulfo-NHS were added and the reaction was stirred for 5 h before adding 32 mg of ConA and left to react overnight at R.T. Finally, samples were centrifuge, washed twice with sterile PBS 1x and suspended in fresh PBS.

### 2.6. Loading with levofloxacin (MSN@LEVO and MSN$_{ConA}$@LEVO)

12 mg of MSN and MSN$_{SATES}$ were collected by centrifugation and suspended in 2 mL of a LEVO solution (2.85 mg/mL) with stirring for 24h. After that MSN@LEVO and MSN$_{SATES}$@LEVO samples were centrifuge, washed once with EtOH and sterile PBS 1x and suspended in 2 mL of fresh PBS.

After that 24 mg of EDC were added to (MSN$_{SATES}$@LEVO) and the mixture was stirred at R.T. for 40 min. Then 14 mg of NHS were added and the reaction was stirred for 5 h before adding 24 mg of ConA and left to react overnight at R.T. Finally, samples were centrifuged, washed twice with sterile PBS 1x and suspended in 1mL of fresh PBS affording MSN@LEVO and MSN$_{ConA}$@LEVO. Both suspensions were divided in two batches, one was dried to performed release experiments and the other one was use for viability experiments. Quantitative determination of LEVO loaded was determined by elemental CHN chemical analyses in a Perkin Elmer 2400 CHN and a LECO CHNS-932 thermo analyzers) and thermogravimetric analysis (TGA) in a Perkin Elmer Pyris Diamond thermobalance. CHN analyses include low analytical ranges of 0.001 – 3.6 mg for carbon, 0.001 – 1.0 mg for hydrogen, 0.001 – 6.0 mg for nitrogen, and 0.001 – 2.0 mg for oxygen.

### 2.7. "*In vial*" cargo release assays

4mg of dried MSN@LEVO and MSN$_{ConA}$@LEVO were suspended in 0.45 mL of PBS 1x and subjected to sonication until a good suspension was achieved. Then, 170 μL of each nanoparticles suspension was filled into a reservoir cap sealed with a dialysis membrane (molecular weight cut-off 12,000 g mol$^{-1}$), allowing released LEVO molecules to pass into the cuvette (which was completely filled with PBS 1x) while the relatively large particles were held back. The experiment was performed in triplicate. The amount of LEVO released was determined by fluorescence measurements ($\lambda_{exc}$ = 292 nm, $\lambda_{em}$ = 494 nm) of the solution recorded on a BioTek Spectrofluorimeter (BioTek Instruments GmbH, Germany). Different calibration lines have been calculated in a concentration range of 12–0.01 μg/mL. To determine the effectiveness of the LEVO dosages released from the different MSNs against bacteria growth, 100 μL of each dosage was inoculated in 900 μL of 10$^8$ bacteria per mL in PBS and incubated overnight. The presence or not of bacteria, as well as their quantification, was determined by counting the colony forming units (CFUs) in agar. In this sense, 10 µL of this solution was seeded onto tryptic soy agar (TSA) and incubated at 37 ºC overnight and subsequent counting. Controls containing bacteria was also performed and the experiments were performed in triplicate.

**2.8. Internalization assays into *E. coli* bacteria biofilm**

*In vitro* targeting assays of LEVO free-nanosystems to the bacteria biofilm has been performed. For this purpose, *E. coli* biofilms were previously developed onto poly-D-lysine coated round cover glasses by suspending the round cover glasses in a bacteria suspension of $10^8$ bacteria per mL during 48 h at 37 ºC under orbital stirring at 100 rpm. In this case, the medium used was 66% THB + 0.2% glucose to promote a robust biofilm formation. After that, the round cover glasses containing the biofilm were placed onto 24 well culture plates (CULTEK) in 1.5 mL of new medium. Then, 0.5 mL of a suspension of red-labelled MSN materials in PBS at final concentration of 5, 10, 50 µg/mL was added. After 90 min of incubation, the glass discs were washed three times with sterile PBS and 3 µL/mL of Syto were added to stain the bacteria live, and 5 µL of calcofluor solution was added to stain the mucopolysaccharides of the biofilm (extracellular matrix) in blue to specifically determine the biofilm formation. Both reactants were incubated for 15 min at RT. Controls containing biofilm bacteria were also performed. Biofilm formation was examined in an Olympus FV1200 confocal microscope and eight photographs (60X magnification) were taken of each sample.

**2.9. Antimicrobial effects against Gram-negative *E. coli* biofilm**

Effectiveness of the LEVO loaded nanosystems against biofilm was also determined. For this purpose, two different assays were performed onto mature *E. coli* biofilm. First, confocal assays where *E. coli* biofilms were previously developed onto poly-D-lysine coated round cover glasses as previously were conducted. Then, 0.5 mL of a suspension of different nanaoparticles in THB at different concentration (5, 10, 50 µg/mL) was added. After 90 min of incubation, the glass discs were washed three times with sterile PBS and 3 µL/mL of Live/ Dead Bacterial Viability Kit (Backlight™) was added. Also, 5 µL/ mL of CALCOFLUOR solution was added to stain the extracellular matrix in blue to specifically determine the biofilm formation. Both reactants were incubated for 15 min at RT. Controls containing biofilm bacteria were also performed. Biofilm formation was examined in an Olympus FV1200 confocal microscope and eight photographs (40X magnification) were taken of each sample. In order to perform quantitative analyses, the surface area covered with live bacteria (green) and extracellular matrix (blue) was calculated using ImageJ software (National Institute of Health, Bethesda, MD) from eight images. All images are representative of three independent experiments. Second, quantitative antibiofilm assays were carried out by calculating the reduction of CFU/mL. Previously, *E. coli* biofilms were seed onto 24-well plates (CULTEK) by incubation of $10^6$ bacteria/mL during 48 h at 37 °C and orbital stirring at 100 rpm. In this case, the medium used was 66% THB + 0.2% glucose to promote a robust biofilm formation. After that, the biofilms were gently washed and 1mL of a suspension of MSN materials in THB at the concentration of 5, 10, 20 μg/mL was added. After 24 h of incubation, the well plates were washed three times with sterile PBS and applied 1-10 min of sonication was applied in a low power bath sonicator (Selecta, Spain) in order to disperse the biofilm in a total volume of 1mL of PBS. Sonicate fluid was then serially diluted 7 times to 1:10 in a final volume of 1 ml. The experiments were performed in triplicate for each dilution of three different experiments. Quantification of the bacteria was carried out in a 1 mL volume using the Drop-plate method [26]. Five 10-μL drops of each dilution were inoculated on Tryptic Soy Agar (TSA) (Sigma Aldrich, USA) plates, which were incubated for 24 h at 37ºC. The mean count of 5 drops of each dilution was made, and then the average counting was calculated.

**2.10 Cell viability in presence of pre-osteoblastic cells**

Cell culture studies were performed with mouse osteoblastic cell line MC3T3-E1 (subclone 4, CRL-2593; American Type Culture Collection, Manassas, VA). First, cells were plated (24-well plates (CULTEK)) at a density of 20,000 cells·cm$^{-2}$ in 1 mL of α-minimum essential medium, containing 10% heat-inactivated fetal bovine serum and 1% penicillin (BioWhittaker Europe)−streptomycin (BioWhittaker Europe) at 37 °C in a humidified atmosphere of 5% $CO_2$, and incubated during 24 h. After that, different concentration of MSNs 5, 10 and 50 μg/mL were placed into each culture. Cell Viability after 24 h of incubation with different MSNs was analyzed. Cell growth was determined by the CellTiter 96AQueous assay (Promega, Madison, WI), a colorimetric method for determining the number of living cells in culture. CellTiter 96 Aqueous one-solution reagent [40 μL, containing 3-(4,5-dimethythizol-2-yl)-5-(3-carboxymethoxyphenyl)-2-(4-sulfophenyl)-2H-tetrazolium salt (MTS) and an electron coupling reagent (phenazine ethosulfate) that allows its combination with MTS to form a stable solution) was added to each well, and plates were incubated for 4 h. The absorbance at 490 nm was then measured in a Unicam UV-500 UV−visible spectrophotometer.

**2.11 Statistics**

All data are expressed as means ± standard deviations of a representative of three independent experiments carried out in triplicate. Statistical analysis was performed using the Statistical Package for the Social Sciences (SPSS) version 19 software. Statistical comparisons were made by analysis of variance (ANOVA). Scheffé test was used for post hoc evaluations of differences among groups. In all of the statistical evaluations, $p < 0.05$ was considered as statistically significant.

**3. Results and Discussion**

**3.1 Preparation and characterization of the nanosystems**

The nanoantibiotic, denoted as $MSN_{ConA}$@LEVO, was synthesized using several steps. (Fig. 2) Briefly, pure silica MSNs were synthetized by the well-known modified Stöber method [27] and externally functionalized by grafting an alkoxysilane bearing carboxylic acid groups, which allows the final anchorage of ConA by reaction with the amine groups present in the protein. LEVO loading was carried out by impregnation method in ethanol [13], and always before ConA grafting to prevent protein denaturation.

With the aim of confirming the chemical grafting of the different functional groups MSN, MSN$_{SATES}$ and MSN$_{ConA}$ were characterized by different techniques and the results compared after each reaction step. By FTIR spectroscopy we were able to follow the nanoparticle functionalization process. The change from a clean spectrum in the 1500- 2000 cm$^{-1}$ range of MSN to the presence of a signal at 1637 cm$^{-1}$ characteristic of the stretching vibration of acid groups was observed in MSN$_{SATES}$ (Fig. 3). Finally in the FTIR spectrum of MSN$_{ConA}$, the presence of amide bonds and NH out of plane bands were clearly seen, confirming the presence of this protein.

ζ-potential measurements in water of these particles show representative changes on the superficial charge with values of -21.5, -33.5 and -25.3 mV for MSN, MSN$_{SATES}$ and MSN$_{ConA}$, respectively. The functionalization degree of the particles was calculated by difference between TGA measures finding 18% of SATES in MSN$_{SATES}$ and 11% of protein in the final system MSN$_{ConA}$. As it would be expected, the pore volume of the MSN suffers a decrease with increasing surface decoration. Thus, the specific surface area changes from 907 in the case of naked material (MSN) to 240 m$^2$/g, in the case of the complete system MSN$_{ConA}$.

Structural characterization by TEM (Fig. 4A-D) shows spherical nanoparticles with an average size of *ca.* 150 nm and a honeycomb mesoporous arrangement before and after functionalization and anchoring of the ConA. Moreover, all samples exhibit typical MCM-41 structure with 2D hexagonal structure, which is also confirmed by XRD studies (Fig. S1). In this case, a small reduction in the intensity of the XRD peaks are observed for MSN$_{SATES}$ and MSN$_{ConA}$, which is attributed to slight order loss that may be ascribed to the partial filling of the mesopore channels by the functionalization agent. In addition, after being stained with 1% of PTA, both samples that contained ConA (MSN$_{ConA}$ and MSN$_{ConA}$@LEVO) show that the protein is cover all external surfaces of the nanoparticles. Keeping the spherical morphology even after loading win LEVO as it can see in Fig. 4D. To acquire information regarding the mean size and stability of the nanosystems in solution, dynamic light scattering (DLS) measurements were recorded (Fig. S2). The measurements performed in H$_2$O MiliQ showed a small increase in the hydrodynamic radius of the particles after the functionalization with SATES and ConA.

**3.2. *In vitro* LEVO release**

The amount of LEVO loaded in nanoparticles was determined by Elemental Chemical Analysis and TGA being of 3.0 and 3.8% in weigh for MSN@LEVO and MSN$_{ConA}$@LEVO respectively, which is comparable with other studies based on silica mesoporous materials [12,28].

The *in vitro* drug release assays from different samples were carried out in PBS at 37ºC and in orbital stirring. Fig. 5 displays *in vitro* release profiles, which are expressed as accumulative drug release as function of the time. The results indicate that both release curves can be adjusted to first order kinetics, being the release rate faster and significantly higher after ConA grafting. Moreover, MSN@LEVO partially retains the loaded drug, being the maximum drug released ca. 30% after 48h, according with previously results [12,29]. On the contrary, MSN$_{ConA}$@LEVO, releases almost the total amount of loaded antibiotic after 5 d of assay. Previous studies have revealed that there is a strong interaction between the LEVO molecules and silica network in MSNs via hydrogen bonding between zwitterionic form of this quinolone antibiotic at pH 7.4 and Si-OH groups of silica nanoparticles.[¡Error! Marcador no definido.] Thus, bare MSN sample partially retains the loaded LEVO, releasing around of 30% after 2 d of test. On the contrary, the presence of ConA protein onto the external surface promotes the interaction with LEVO molecules, which provokes drug departure from the mesopores resulting in a faster antibiotic release.

Besides, the biological activity of each antibiotic doses at the different tested times was also evaluated by incubation with *E. coli* suspensions ($10^8$ bacteria/mL) and subsequent counting of colony forming units (CFUs). An antimicrobial efficacy after 2 d and 5 d for MSN@LEVO and MSN$_{ConA}$@LEVO respectively was observed (Fig. S3), being in good agreement with kinetics studies.

In general, the drug release kinetics from mesoporous matrices are governed, primarily by drug diffusion processes throughout the matrix. Such drug diffusion processes are fitted, generally, to the Higuchi model. However, our results suggest that in addition to the drug diffusion process throughout the mesoporous matrix, a new component is governing the drug release kinetics. Specifically, this new component refers to the silica matrix-LEVO interactions, as it has been previously reported for other silica matrix.

**3.3. *In vitro* internalization assays (targeting effect)**

After demonstrating the antimicrobial capacity of the nanoantibiotic, the next step consisted in evaluating its bacterial biofilm targeting efficacy. For this purpose, a preformed *E. coli* biofilm was incubated with different concentrations (5, 10 and 50 µg/mL) of nanoparticles suspensions. To solely evaluating the targeting effect free-LEVO nanosystems were used and confocal microscopy studies were conducted at different depths. To visualize the nanoparticles, they were labelled in red with rhodamineB (RhB). Fig. 6 shows the internalization study of pristine MSN and MSN$_{ConA}$ in a preformed *E. coli* biofilm after 90 min of incubation and 50 µg/mL of nanoparticles, chosen as a representative concentration. 3D confocal reconstruction evidences a typical biofilm structure, live bacteria (green) covered by a protective polysaccharide matrix (blue), where MSN are localized onto its surface (see white arrows). These results are consistent with what was published in other studies. In 2014 Forier *et al.* investigated the effect of nanoparticle size in biofilm penetration, indicating that the cutoff size for optimal penetration was around 130 nm [30]. Specifically for MSN, D. L. Slomberg et al. synthesized two sizes (14 and 150 nm) of nanoparticles and studied their diffusion into *P. aeruginosa* biofilm. They observed that diffusion was also size dependent and, although both MSN penetrated into the biofilm, the process was higher and faster for the smaller ones [31]. Therefore, it was expected that our pristine MSN, whose size (150 nm) is within the limit described, would be mainly retained on the surface of the biofilm. On the contrary, what is surprising is that MSN$_{ConA}$ are able to penetrate the biofilm and place at different depth levels along the z-axis, suggesting that the effect of ConA is so powerful that force the internalization of nanoparticles whose size is theoretically inadequate. It is also worth of mention that the MSN$_{ConA}$ internalization degree is dose-dependent, i.e. the greater the concentration of added nanoparticles the higher the amount of nanoparticles penetrating the biofilm (Fig. S4). The MSN$_{ConA}$, with a $\zeta$-potential of -25 mV, could have strong electrostatic affinity by the polysaccharide of the biofilm matrix [32]. Moreover, this targeting agent has shown a selective affinity to α-mannopyranosyl and α-glucopyranosyl residues presents in the extracellular polysaccharide matrix, which could explain the internalization mechanism of MSN$_{ConA}$ system into bacterial biofilm [18].

### 3.4. Antimicrobial effects against Gram-negative *E. coli* biofilm

Once the nanosystem has reached the biofilm the anti-bacterial activity is determined by the physicochemical properties of the entrapped antimicrobial inside of the nanosystem. Among others LEVO, a synthetic fluoroquinolone antibacterial agent that inhibits the supercoiling activity of bacterial DNA gyrase, halting DNA replication, is a broad spectrum antibiotic that provides clinical and bacteriological efficacy in a range of infections [33]. However, previous studies [12] have proved its inefficiency once the biofilm has been formed. To solve that, LEVO agent has been incorporated to the mesostructured arrangement of these targeted-nanosystems to be released within biofilm. For this purpose, a preformed *E. coli* biofilm was incubated with different concentrations (5, 10 and 50 µg/mL) of loaded nanoparticles (Fig. S5), being 10 µg/mL the most optimum anti-biofilm concentration.

Fig.7 shows the *in vitro* antimicrobial efficacy against preformed *E. coli* biofilms after 90 min of incubation with a suspension of 10 µg/mL of MSN, MSN@LEVO, $MSN_{ConA}$ and $MSN_{ConA}$@LEVO. The confocal microscopy image corresponding to control (Fig. 7C) shows a typical biofilm formed mainly by a mantle of live bacteria (green) with some dead bacteria (red) isolated and covered with a polysaccharide matrix represented in blue. The results show that samples treated with MSN present a small reduction of the biofilm (Fig. 7A). After treatment with MSN@LEVO (Fig. 7B), the reduction is more visible, appearing gaps in its surface, probably by a more superficial action of the antibiotic since the particles practically did not penetrate into the biofilm, but still showing a large amount of live bacteria. This reduction, in terms of percentage, corresponds to 50% of live bacteria and 70% for covered biofilm for bare MSN@LEVO, showing its antimicrobial inefficiency. On the contrary, this scenario totally changes after treatment with $MSN_{ConA}$ (Fig. 7D) where a significant reduction of the biofilm is observed, even in the absence of the LEVO. Undoubtedly, this increase in antimicrobial capacity must be due to the presence of ConA. A study of the effect produced in the biofilm by $MSN_{ConA}$ sample shows a notable reduction of biofilm of 65% (green scattered) and 75% (blue scattered) (see Fig.7F). More experiments are necessary to elucidate the mechanism of ConA toxicity in the biofilm. However, we hypothesized that it could be similar to that observed for some other lectins [34-36]. Finally, figure 7E shows that the biofilm treated with $MSN_{ConA}$@LEVO,

suffered a complete eradication, appearing only a few scattered of bacteria (mostly dead/red). A deep analyses surface, calculated using ImageJ software, showed a reduction of almost 99.9% of live bacteria and almost 100% of blue covered biofilm for the targeted-nanosystems. These results evidence that the incorporation of ConA as biofilm targeting agent onto the MSN surface platform triggers the complete biofilm destruction in combination with antibiotic. In this sense, due to the penetration of these nanosystems into the biofilm and the release of the antibiotic inside of it, high antimicrobial efficacy is produced. By the contrary, the treatment with bare MSN@LEVO provokes a reduction in the biofilm area but the effect was significantly less effective due to the lack of penetration into the biofilm. In addition, to determine the antibiofilm effect of MSN samples and confirm the effective dosage bacteria contained into biofilm after treatment were counted by Drop-plate method. Fig. 8 displays the reduction percentage of biofilm in terms of CFU per mL with respect to control after treatment. The obtained results show a notable reduction of 97.8 and 100 % for 10 and 20 mg/mL for MSN$_{ConA}$@LEVO, respectively in agreement with confocal microscopy assays (Fig.7D). Note that the dose at 10 μg/mL shows a bacterial concentration of less than $10^2$ bacteria/mL, which could be considered as an effective treatment [37].

### 3.5. Biocompatibility assays

The use of these nanocarriers in infection treatment for clinical applications requires that the designed material present excellent biocompatibility and absence of cytotoxicity. MSN is a biocompatible material that exhibits low toxicity and lack of immunogenicity and is degraded into nontoxic compounds (mainly silicic acid) in relatively short time periods [38]. Despite its lack of toxicity, the surface modification of this MSN could provoke the appearance of toxicity due to enhanced uptake within the cells. To evaluate toxicity, MC3T3-E1 cells were incubated with different amount of the MSNs in cell culture medium for 24 h. After this time, cell viability was determined via the standard cell viability test by MTS reduction. The results showed that none of these empty materials exhibited cytotoxicity in either cell line (Fig. 9) in agreement with previous results [19].

## 4. Conclusions

In this work a new class of targeting antimicrobial device, based on mesoporous silica nanoparticles (MSNs) decorated with ConA and loaded with LEVO as antibiotic has been developed. The covalently grafting of ConA to MSNs (MSN$_{ConA}$) allows an effective penetration in Gram-negative bacteria biofilm, which increases the antimicrobial efficacy of LEVO hosted in the mesopores. These findings demonstrate that the synergistic combination of biofilm internalization and antimicrobial agents into a unique nanosystem provokes a remarkable antimicrobial effect against bacterial biofilm. This nanocarrier is a promising alternative to the current available treatments for the management of infection. The next step is to determine its clinical relevance by *in vivo* models in wounds or after implant and prosthesis application.

**Apendix A. Supplementary data**

Supplementary data associated with this article can be found in the online version.

**Acknowledgements**

This work was supported by European Research Council, ERC-2015-AdG (VERDI), Proposal No. 694160.

**Figure captions**

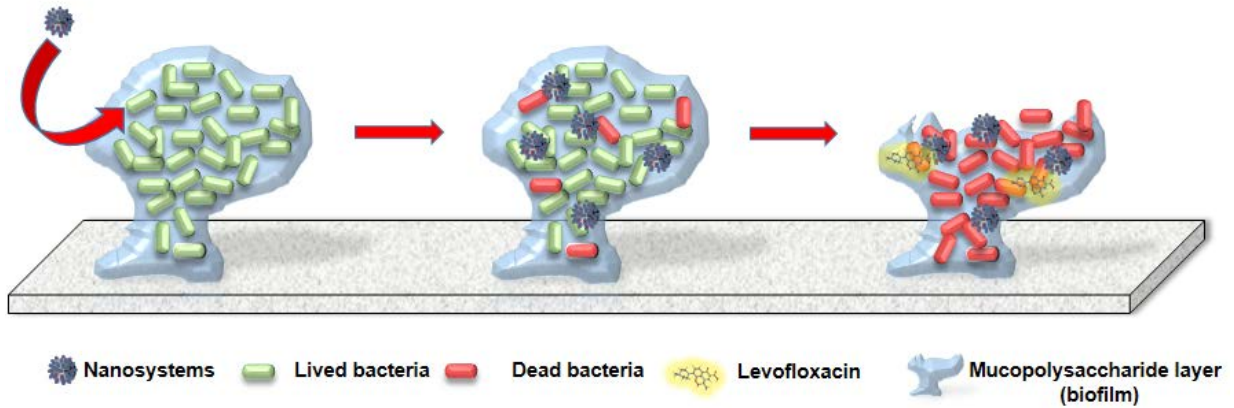

**Fig. 1.** Schematic illustration of the aim of this work.

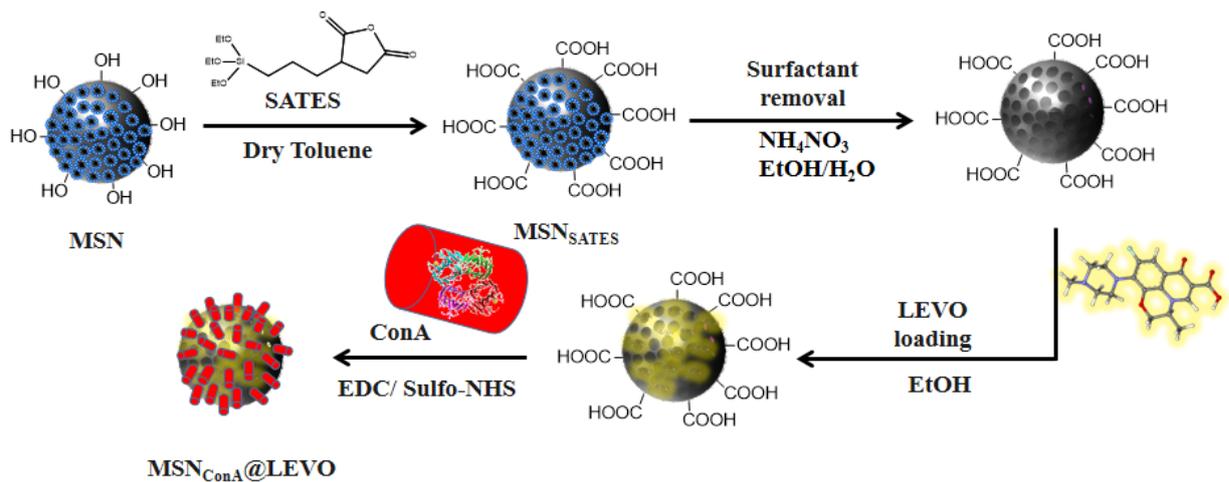

**Fig. 2.** Schematic depiction of the synthesis procedure used to develop our nanoantibiotic: (i) Functionalization with carboxylic groups (MSN$_{SATES}$); (ii) surfactant extraction; (iii) load with LEVO; (iv) anchoring of ConA into external surface of nanoparticles (MSN$_{ConA}$@LEVO).

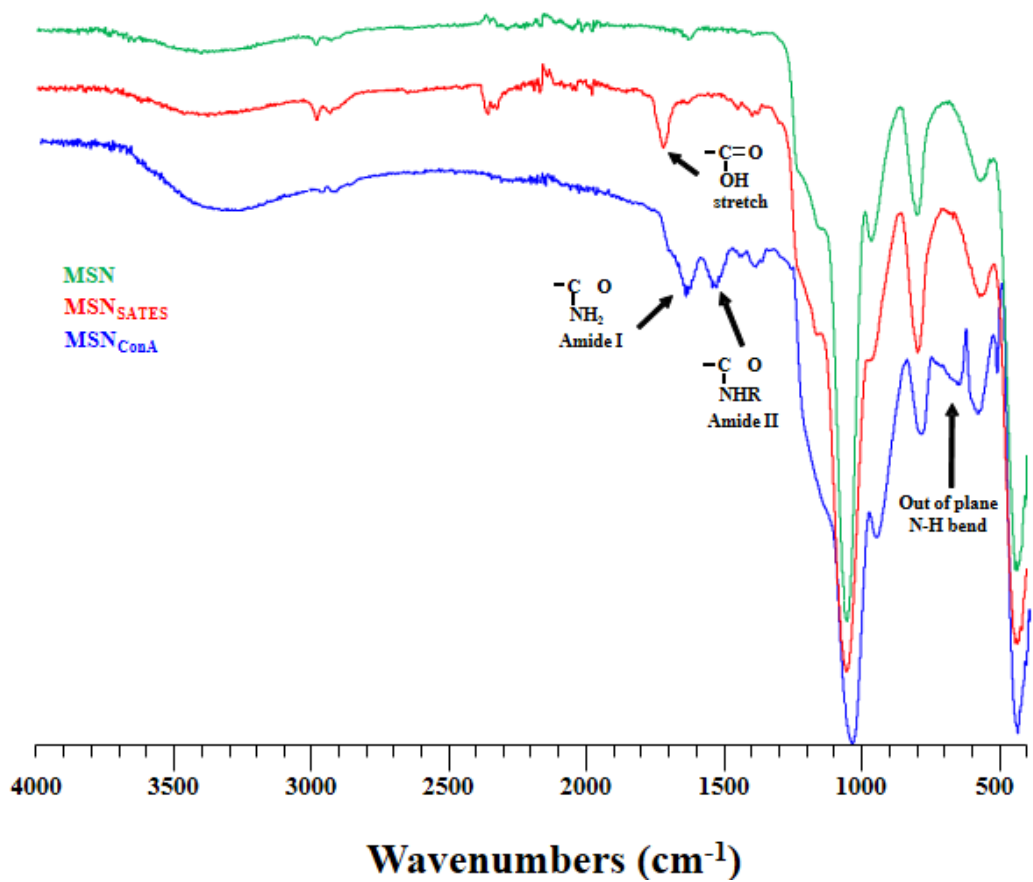

**Fig. 3.** FTIR spectra of MSN, MSN$_{SATES}$ and MSN$_{ConA}$, confirming the effectiveness of functionalization process.

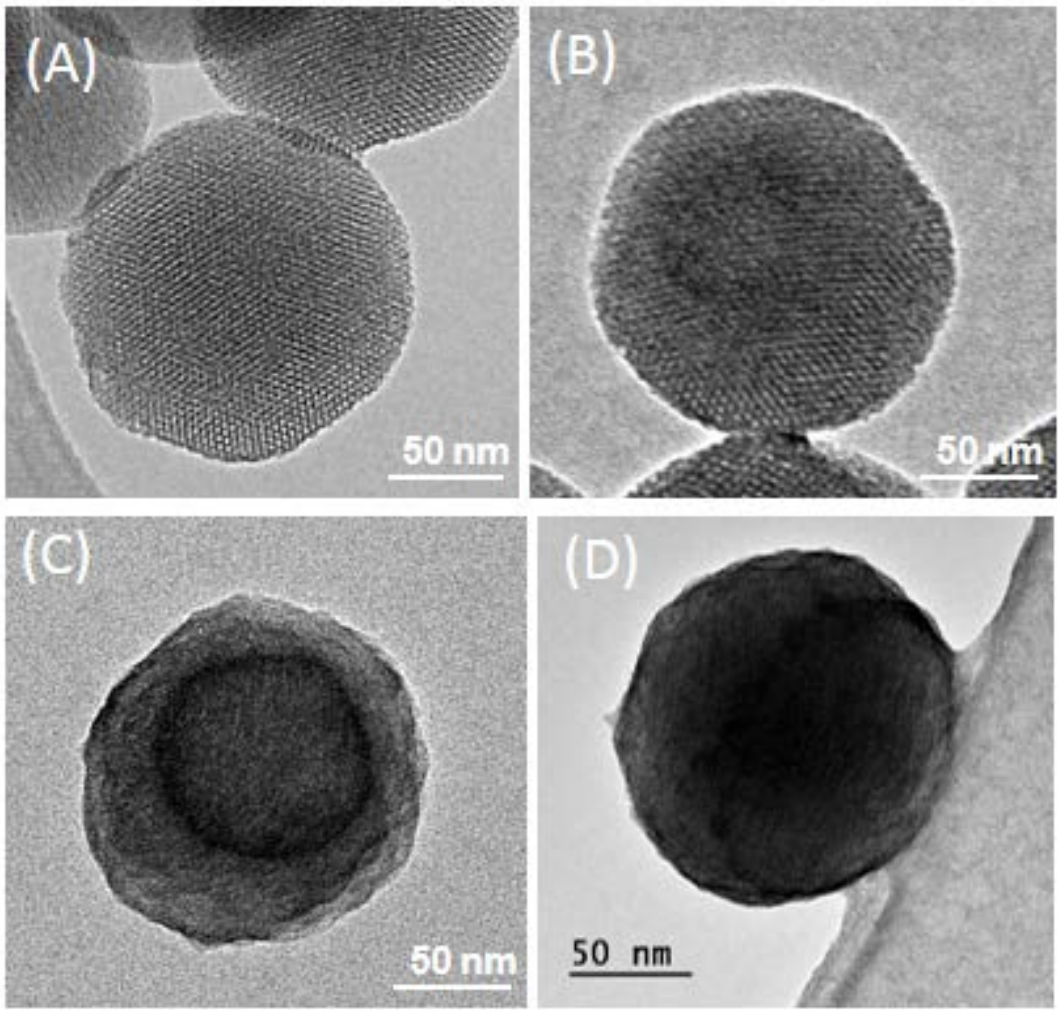

**Fig. 4.** TEM images of (A) pristine MSN, (B) MSN$_{SATES}$, (C) MSN$_{ConA}$ and (D) MSN$_{ConA}$@LEVO, respectively. The samples after functionalized with the organic compound were stained with 1% PTA in order to visualize by this technique (images C and D).

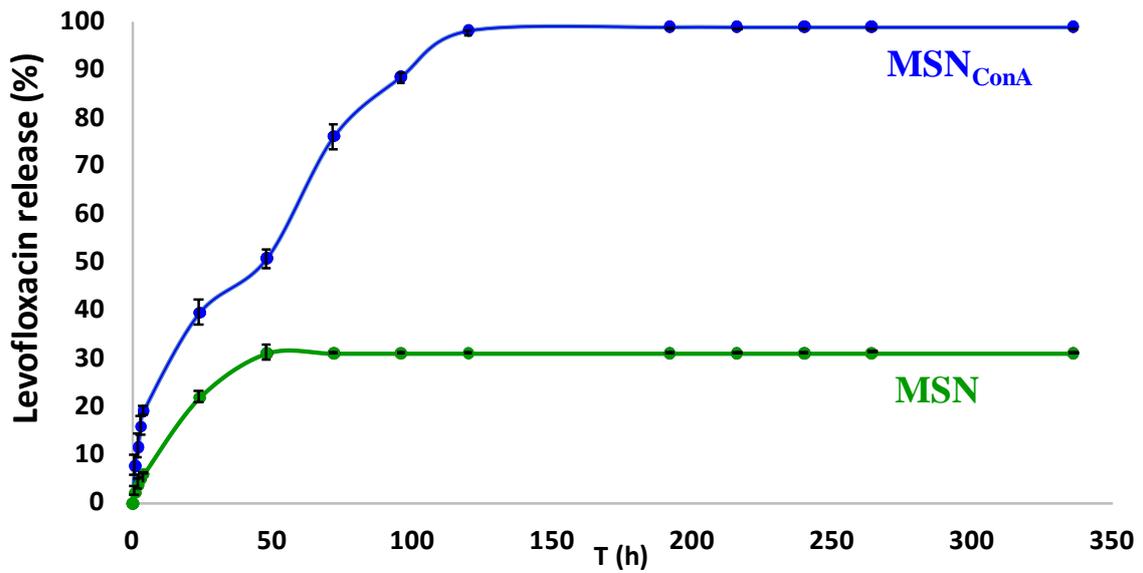

**Fig. 5.** *In vitro* LEVO release profiles (n = 3) from MSN<sub>ConA</sub> and MSN in PBS at 37ºC and orbital stirring, The amount of LEVO released was determined by fluorescence measurements of the solutions, ($\lambda$ex = 292 nm, $\lambda$em = 494 nm) and expressed as accumulative drug release as function of the time.

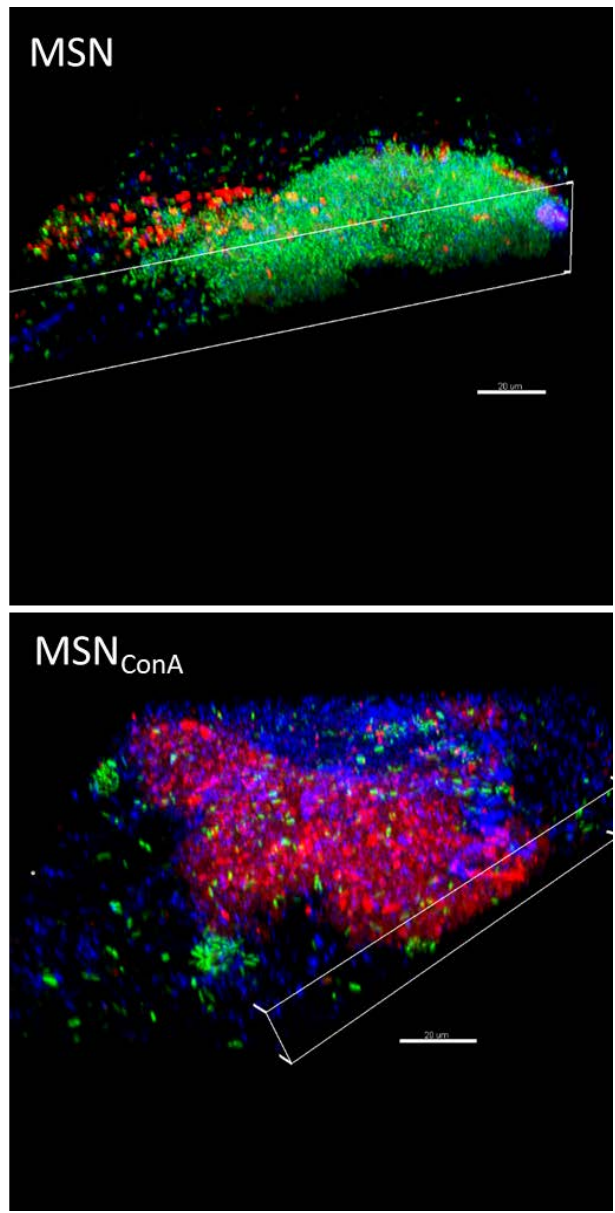

**Fig. 6.** Internalization study by confocal microscopy study of red-labelled pristine MSN and MSN$_{ConA}$ in preformed *E. coli* biofilms after 90 min of incubation and 50 μg/mL of nanoparticles. 3D confocal reconstruction shows that bare MSN are localized onto the biofilm surface whereas MSN$_{ConA}$ penetrate the biofilm and are placed at different depth levels. Live bacteria are stained in green (SYTO), nanoparticles in red (RhB) and the extracellular polysaccharide biofilm matrix in blue (CALCOFLUOR). Scale bars, 20 μm.

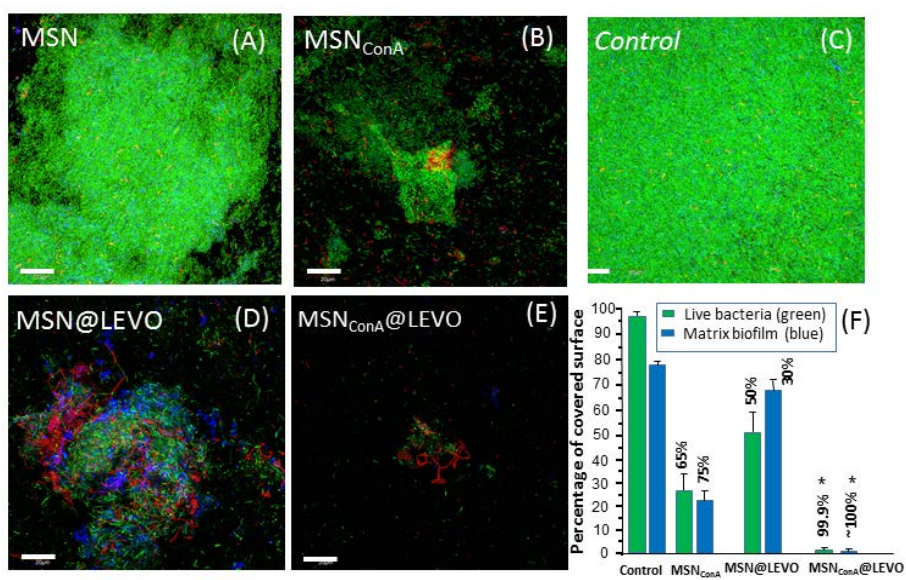

**Fig. 7.** Confocal microscopy study of the antimicrobial activity of the LEVO loaded MSNs materials onto Gram-negative *E. coli* biofilm. The confocal images show (A) the biofilm preformed onto covered glass-disk without treatment (control) and after 90 min of incubation with (B) MSN@LEVO and (C) MSN$_{ConA}$@LEVO, respectively. Live bacteria are stained in green, dead bacteria in red and the protective matrix biofilm in blue. Scale bars, 20 µm. (D) Histogram representing the percentage covered surface in green and blue from eight confocal images and calculated by ImageJ software (National Institute of Health, Bethesda, MD) after treatment with 10 μg/mL of different nanoparticles. The numerical data represent the percentage of reduction in each case with respect to control (in absence of any nanoparticle treatment). The experiments were performed in triplicate. *p <0.05 *vs* corresponding MSN@LEVO and MSN$_{ConA}$ (ANOVA).

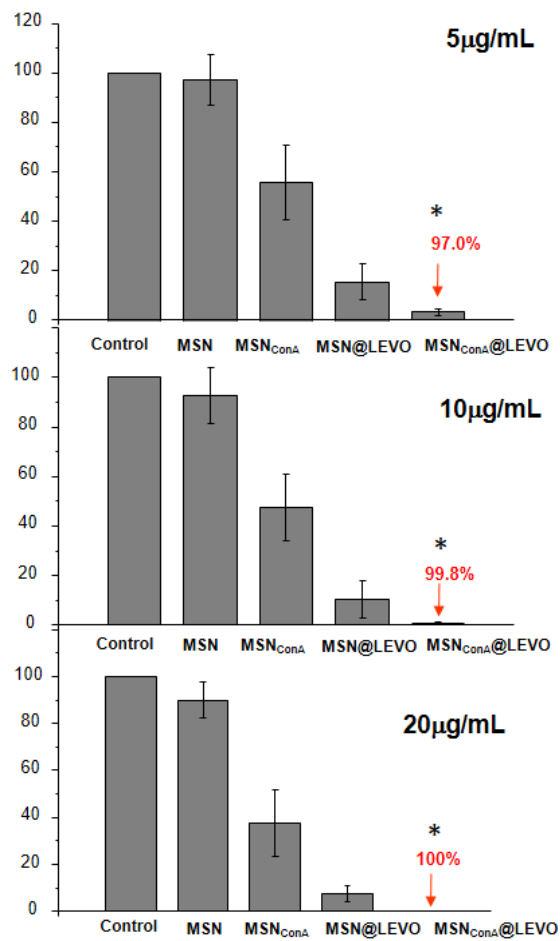

**Fig. 8** Effect on *E.coli* biofilm of the samples at different concentration after 24 h. It is represented the reduction percentage with respect to control in absence of nanoparticles. *p <0.05 *vs* corresponding MSN@LEVO (ANOVA).

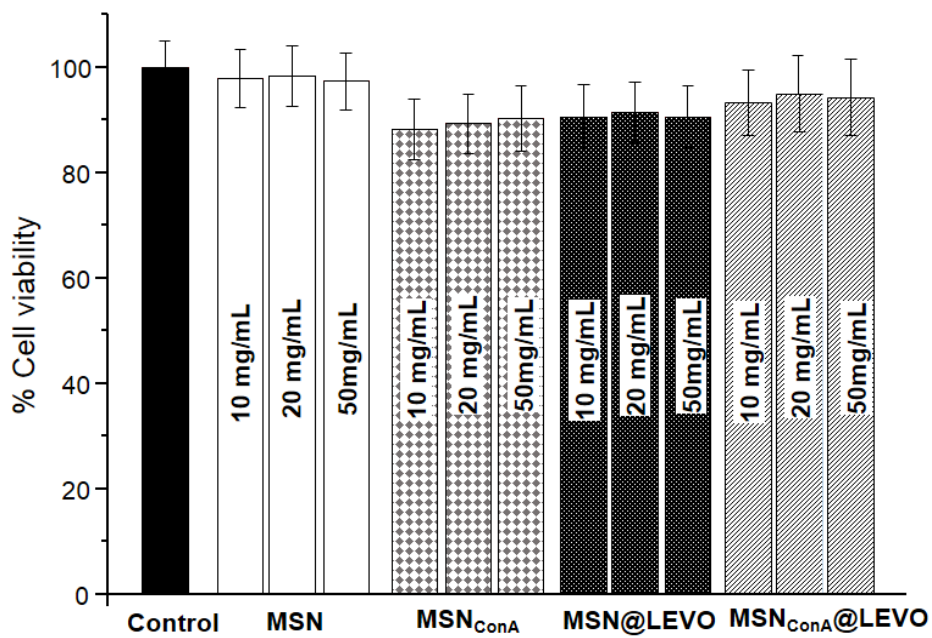

**Fig. 9.** Cell viability studies of the samples at different concentration for MC3T3-E1 cell line and 24 h of exposure time. *p <0.05 *vs* corresponding control without nanoparticles (ANOVA).